\documentclass[twocolumn,showpacs,aps,superscriptaddress,psfig]{revtex4}
\usepackage{graphicx}

\begin{document}\draft

\title{Clustering of floaters by waves}

\author{P. Denissenko$^{1}$, G. Falkovich$^{\dag}$ and
S. Lukaschuk}
\address{Fluid Dynamics Laboratory, University of Hull, HU6 7RX,
UK\\
$^{\dag}$Physics of Complex Systems, Weizmann Institute of Science,
Rehovot 76100, Israel}
%\date{\today}
\begin{abstract}
We study experimentally how waves affect distribution of particles
that float on a water surface. We show that clustering of small
particles in a standing wave is a nonlinear effect with the
clustering time decreasing as the square of the wave amplitude. In a
set of random waves, we show that small floaters concentrate on a
multi-fractal set.
\end{abstract}\pacs{47.27.Qb, 05.40.-a}

\maketitle Even for incompressible liquids, surface flows are
generally compressible and can concentrate pollutants and
floaters.
%(a simple example given by the surface flow due to two parallel
%counter-rotating underwater turbines).
Spatially smooth random
flows can be characterized by the Lyapunov exponents whose sum is
the asymptotic in time rate of volume change in the Lagrangian
frame (co-moving with the fluid element). Since contracting
regions contain more fluid particles and thus have more
statistical weight than expanding ones, the rate is generally
negative in a smooth flow (for volume in the phase space, this is
a particular case of the second law of thermodynamics)
\cite{Ruelle97,BFF,FF,FGV}. As a result, density concentrates on a
fractal (Sinai-Ruelle-Bowen) measure in a random compressible flow
\cite{Sinai,BR75, Ruelle, Dorfman}. Indeed, it has been observed
experimentally that random currents concentrate surface density on
a fractal set \cite{Som1,Som,Nam,CB,CB1}. Moreover, recent theory
predicts that the measure must actually be multi-fractal i.e. the
scaling exponents of the density moments do not grow linearly with
the order of the moment \cite{BFF,BGH,Dav}.

Here we study the effect of clustering by waves on the water
surface. In a single-mode standing wave, fluid surface expands and
contracts periodically. Only in a set of random waves, one may find
regions where contractions accumulate and lead to the growth of
concentration. This is true, yet for potential waves the respective
rate of clustering of the points on the water surface appears only
in the sixth order in wave amplitudes \cite{BFS,Masha}. Since wave
amplitudes are typically much less than the wavelengths (otherwise,
waves break), such a rate is usually so small as to be unobservable.
For example, for waves with periods in seconds and the (pretty
large) ratio of the amplitude to wavelength $0.1$, the clustering
time is in weeks. However, even small particles can move relative to
the fluid. The physical mechanism that causes drift of floaters
relative to water surface is the capillarity which breaks
Archimedes' law and makes a floater inertial (i.e. lighter or
heavier than the displaced liquid). As a result, the floaters
cluster already in a standing wave (either in the nodes or in the
antinodes  depending on the sign of the capillary force)
--- brief report of the discovery of this phenomenon has been
published in \cite{Nat}. The theory of particle motion in a
standing surface wave, presented in the Supplement to \cite{Nat},
predicts that the drift must appear in the second order in wave
amplitudes. The first part of our experimental results described
here shows that this is indeed the case.

%Here we show experimentally that this is indeed the case.

%Inertial effects bring fast clustering as we shall show below. We
%also demonstrate experimentally  that floaters concentrate on a
%multi-fractal set in a flow of random waves.
We measured clustering time  for small hydrophilic hollow spheres
with the average size 30$\,\mu m$ and density $0.6$ g/cm$^3$.
Particles were sift from the dry powder of glass bubbles (S60HS,
3M), separated by flotation in acetone (density 0.78 g/cm$^3$) and
washed in clean water. The particles spread readily over flat water
surface and do not form stable clusters. This is possibly caused by
double layer repulsion, which compensates an attraction due to the
surface tension.

%The water suspension of particles
%was added just before sealing and final level adjustment.
The experimental set-up is shown in Figure 1. Surface waves are
generated in a rectangular cell (C) (horizontal size 9.6 x 58.3 mm,
and depth 10 mm) through the parametric instability \cite{Far}. The
cell is filled with purified water (resistivity 18 MOhm$\cdot$cm) up
to the edge of the lateral walls to eliminate the meniscus effect -
"brim-full" boundary conditions. The cell is sealed and mechanically
coupled to the electromagnetic shaker V (V20, Gearing and Watson
Electronics Ltd), whose vertical oscillation amplitude and frequency
is controlled by digital synthesizer (Wavetek 81). The oscillation
amplitude is measured by iMEMS accelerometer ADXL150 (Analog Device)
attached to the moving frame.
\begin{figure}
\includegraphics[width=7.4cm,height=4.6cm]{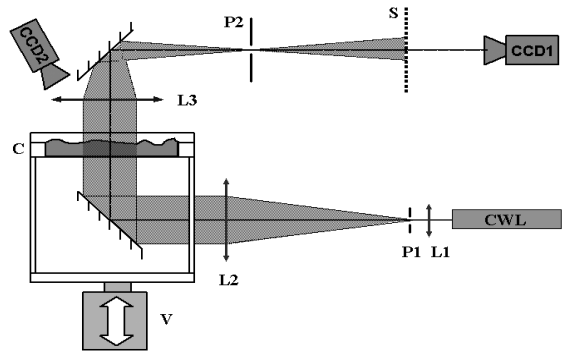}
\caption{Experimental setup.}
\end{figure}%\vskip 0.1truecm
The cell is illuminated from below by the expanded collimated beam
(pin-hole P1, lenses L1-L2) from the continuous wave laser (CWL). A
spatial filter, the lens L3 with 0.1 mm pin-hole P2 at focal
distance F=250 mm, rejects all refracted light and forms an image of
the anti-nodes on the screen S. Two high-resolution cameras CCD1 and
CCD2 (2048 x 2048 pixels) are controlled through the Dantec PIV
system. The shutters of both cameras  are synchronized in phase with
the shaker oscillation. The camera CCD1 collects the images of
anti-nodes. Its shatter is open for a time equal to the one period
of the parametric wave. The camera CCD2 collects the light scattered
by the particles on the surface. It is positioned off axis to avoid
straight laser light and its shutter is opened for a shorter time
($\sim$1 ms) to prevent smearing of particle in the images. The CCD2
shutter is opening at the phase when the liquid surface is nearly
flat. This allows to keep CCD2 at a minimal angle to the system's
optical axis. The optical axis of CCD2 is perpendicular to the cell
long axis.

The  measurements of the clustering time  as a function of the wave
amplitude were performed as follows. For a given cell geometry and
chosen frequency, we determined  the parametric instability
threshold, an oscillation amplitude $A_c$. This procedure was
similar to that described in \cite{SD}. Each experimental run  has
been started from mixing: the shaker amplitude was kept at
$A\sim5A_c$ for a couple minutes and then lowered to $A\sim0.9A_c$.
Next, a desired amplitude of vibration $A_i>A_c$ is set and the
acquisition of the images from both CCDs started. The unstable
parametric wave appears after a time delay with an amplitude growing
up to a stationary value proportional $(A-A_c)^2$. A set of
collected images always starts from the moment when there are
neither waves nor particle motion and ends when a new stationary
state reached with the developed wave and clustered particles. After
the parametric wave appears, the homogeneous area in CCD1 images is
replaced by a network of lines corresponding to the wave anti-nodes
(since the water surface curved by the wave serves as a lens). The
local line width decreases as the wave amplitude increases, and the
maximum of intensity is constant across the line. So the variance of
the light intensity averaged over an image area can be chosen as a
characteristic of the wave amplitude. The variance of intensity
amplitude measured from the particle images (CCD2) represents the
amplitude of the growing inhomogeneity in particle concentration. A
number of frames collected for each camera is 100 and the frame rate
is adjusted using preliminary test runs.

Figure 2a shows the original images of anti-nodes and particle
clusters. Figures 2 b,c,d present the results of image processing -
the variances of light intensities from CCD1 and CCD2 versus time
showing how the particle clustering and the wave develop. The bottom
curve 2d shows the wave amplitude taken from CCD1 while the upper
two curves, 2b and 2c, present respectively the longitudinal and
lateral variances of particle concentration taken from the variances
of light intensity measured by CCD2. This allows to observe
clustering along each wave vector of the standing wave separately.

\begin{figure}
\includegraphics[width=8.4cm,height=6.8cm]{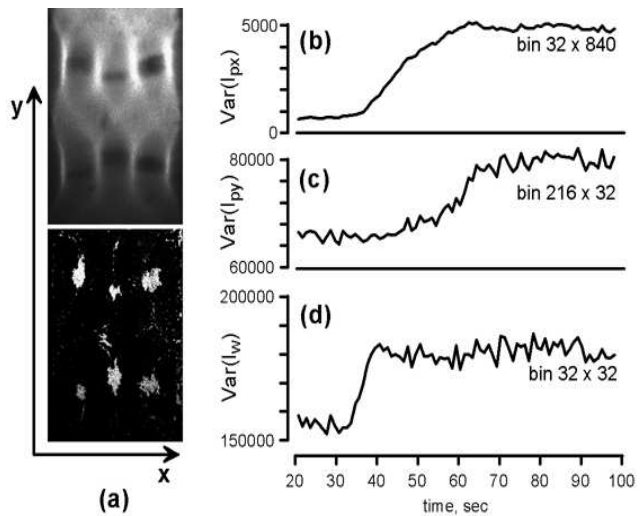}
\caption{Visualization of the standing wave and particles
clustering at the nodes (a), the variance of light intensities as
a function of time showing the growth of lateral (b) and
longitudinal (c) inhomogeneities in the concentration of particles
and the growth of wave amplitude (d)}
\end{figure}%\vskip 0.2truecm

The time delay between the stabilization of the wave amplitude and
the saturation of concentration inhomogeneity is used as a
characteristic time of clustering. The inverse clustering time is
plotted versus squared wave amplitude in Figure 3. The averaged
surface wave amplitude $<A^2>$ was determined from the sizes $S_x,
S_y$ of the image producing by refracted light in the focal plane P2
of the lens L3. For the refractive index of water 1.33 and small
wave amplitudes, the angles $S_y/F$ and $S_x/F$ are equal to one
third of the maximum surface inclination. It follows from Fig. 3
that within 10\% the inverse clustering time is proportional to the
square of amplitude, as predicted by the theory \cite{Nat}.
%The averaged wave amplitude has been estimated measuring a size of
%the spot at the focal plane.
\begin{figure}
\includegraphics[width=7.4cm,height=5.8cm]{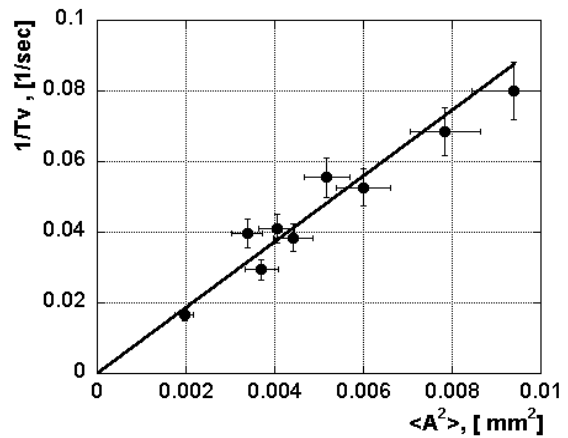}
\caption{The inverse characteristic time of (lateral) clustering
as a function of the squared wave amplitude.}
\end{figure}%\vskip 0.2truecm

Quasi-linear standing waves exist only at small amplitudes of shaker
vibrations. Increasing the amplitude one observes more and more
complicated patterns, from spatio-temporal chaos to developed wave
turbulence, see e.g. \cite{SD,gollub,ZLF,Put,Cap}. Random
compressible flows are generally expected to mix and disperse at the
scales larger than the correlation scale of the velocity gradients
and to produce very inhomogeneous distribution at smaller scales see
\cite{HH,FGV,BFF,BGH,BFS,WG,BM,Bruno} for theory and
\cite{gorlum,Som1,Som,Nam,alstrom,CB} for experiments. Here we show
experimentally that this is also true for a set of random surface
waves (for what follows, it is actually enough if particle motion
corresponds to so-called Lagrangian chaos). Let us briefly describe
relevant mathematical quantities of interest. Consider the number of
particles inside the circle of the radius $r$ around the point ${\bf
x}$: $n_r({\bf x})=\int_{|{\bf r}'-{\bf x}|<r} n({\bf r}')\,d{\bf
r}'$. One asks how the statistics of the random field $n_r({\bf x})$
changes with the scale of resolution $r$. That can be characterized
by the scaling exponents, $\zeta_m$, of the moments: $\langle
n_r^m\rangle\propto r^{\zeta_m}$. Note that $\zeta_0=0$ and
$\zeta_1=2$. When the distribution is uniform on a surface, one
expects $\zeta_m=2m$. When this equality breaks for some $m$, one
usually calls the distribution fractal. First, the fractal
(information) dimension for a random surface flow has been measured
by Sommerer and Ott, who found non-integer $d=d\zeta_m/dm|_{m=0}$
\cite{Som1}. Then, the scaling of the second moment has been found
and related to the correlation dimension (again non-integer)
\cite{Som,Nam}. Therefore, fractality of the distribution has been
established in \cite{Som1,Som,Nam}. To the best of our knowledge,
different dimensions have not been compared for the same flow (if
found different, that would give a direct proof of multifractality).

On the other hand, a theory recently developed for a
short-correlated compressible flow gives the set of exponents
\cite{BFF,BGH} $\zeta_m$ which depend nonlinearly on $m$ (for
comparison, note that those theoretical formulas give the Lagrangian
exponents which in our notations are $\zeta_{n+1}-2$). Such
nonlinear dependence corresponds to a multifractal distribution.
Multifractality of the measure predicted in \cite{BFF,BGH}  means
that the statistics is not scale-invariant: strong fluctuations of
particle concentration are getting more probable as one goes to
smaller scales (increases resolution).

In the second part of the experiment we measured concentration
moments and scaling exponents for the suspension of small
hydrophilic particles mixed by a surface wave turbulence (at the
driving amplitude $\sim2A_c$). Note that at such an amplitude, it is
not yet developed turbulence but rather few modes that interact
nonlinearly and provide for the Lagrangian chaos. The experiment has
been done for a set of oscillation frequencies from 30 to 220 Hz and
amplitudes $1.8-2.5A_c$. We reproduce here a typical result for the
parametric wave with the frequency 32 Hz, wavelength about 7 mm at
the oscillation amplitude 198$\mu m \simeq 2A_c$. A snapshot of the
floaters distribution for this set is shown in Figure 4.
\begin{figure}
\includegraphics[width=8.4cm,height=7.64cm]{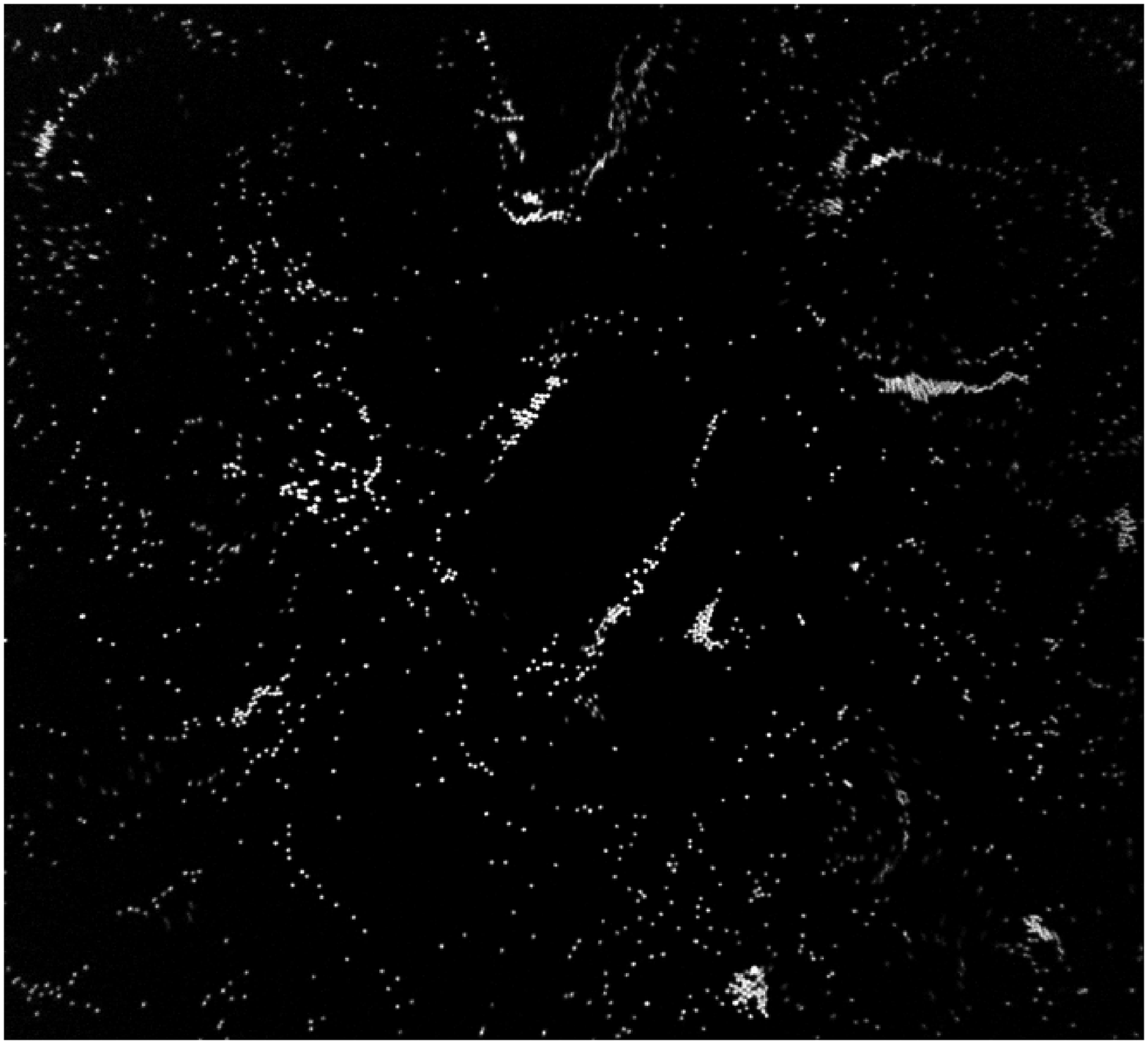}
\caption{The image of particle distribution in random waves (20 x 20
mm). The number of particles 920.}
\end{figure}\vskip 0.2truecm
We checked several approaches to quantify the particle concentration
and found that the most reliable algorithm should base on
recognition and counting the individual particles. We developed such
an algorithm and implemented it for 95 $\mu m$ fluorescent
microspheres (No. Duke Scientific Corp.). The particle density is
1.05 g/cm$^3$. To make them floating we used 20\% salt (NaCl)
solution. Fluorescence method greatly improves the image contrast,
eliminates the problem of spurious refractions, and allows
positioning the CCD2 camera with an optical filter on the system
optical axis. In this part of the experiment we used the cell with
the horizontal span 50x50 mm and the depth 10 mm. The size of
observation area is about 30x30 mm, the pixel size 15 $\mu m$ and
mean particle diameter corresponds to 6-7 pixels.

Images from the experiment with chaotic clustering were
preprocessed. %Non-homogeneous illumination was compensated dividing
%each image by matrix of coefficients recorded using diffusive glass.
A background noise was subtracted using individual threshold for
each frame equal to the mean intensity plus 3 standard deviations.
The resulted images were smoothed by low-passed 5x5 pixels filter.
The particle coordinates were determined maximizing a correlation of
3x3 matrix ($75\times75\,\mu m$). The method was validated by
comparison the number of particles with that estimated on the stage
of emulsion preparation. The particle detection in the dense
clusters was verified by direct visual inspection of images.

\begin{figure}
\includegraphics[width=8.4cm,height=6.84cm]{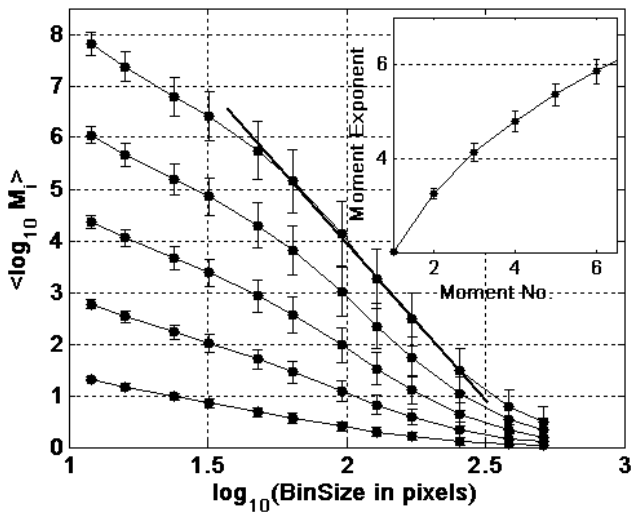}
\caption{Moments of concentration (2,3,4,5 and 6th) versus the
scale of coarse-graining. Inset: scaling exponents of the moments
of particle number versus moment number.}
\end{figure}\vskip 0.2truecm

Up to 1000 images with particle distributions were recorded  at
sampling rate 4 sec.
%A typical standard deviation of the number of
%particles in the open area 30x30 mm averaged over 1000 images is
%less than 10\%.
The first six moments of the coarse-grained concentration,
$N_m=\langle n_r^m\rangle r^{-2m}$, are shown in Figure 5 versus the
scale of averaging (bin size $r$). We see that indeed the moments
with $m>1$ grow when $r$ decreases below the wavelength of the
parametrically excited mode. We see that this growth slows down when
$r$ decreases below $r=50$ pixels ($\lg_10(50)\simeq 1.7$). This is
possibly due to the dense clusters where the finite particle size,
short range repulsion, and the particle back reaction on the flow
are important. An additional reason may be an insufficient
representation of dense regions by the finite number of particles.
On this log-log plot the straight lines correspond to the power
laws. The scaling exponents for the interval $50<r<300$ pixels are
shown in the inset. The nonlinearity of the dependence of $\zeta$ on
$m$ is the first experimental demonstration of multifractality in
the distribution of particles.

Another interesting aspect of particle distribution is related to
the inertia which may cause particle paths to intersect. This
phenomenon was predicted in \cite{FFS} and called sling effect, it
must lead to appearance of caustics in particle distribution
\cite{WM}. At weak inertia (like in our system), caustics are
(exponentially) rare \cite{FFS,WM} yet we likely see one at the
center of Fig.~4. Indeed, as argued in \cite{FFS}, breakdowns of
particle flow are mostly one-dimensional so that caustics must look
locally as two parallel straight lines. At higher inertia,
breakdowns provide for extra mixing that makes the sum of Lyapunov
exponents positive and measure smooth (rather than multi-fractal)
\cite{Bec,MW1}. Statistical signatures of co-existence of caustics
and multi-fractal distribution needs further studies.

We believe that the new effect of clustering by waves is of
fundamental interest in physics and may be of practical use for
particle separation, cleaning of liquid surfaces, better
understanding of environmental phenomena associated with the wave
transport. Chaotic motion of floaters produces the multi-fractal
distribution and caustics and is a good model to study behavior of
inertial particles in random flows important for water droplets in
clouds, fuel droplets in internal combustion engines, formation of
planetesimals etc.

The work is supported by the grants of Royal Society, Israel
Science Foundation and European network. We thank V. Steinberg for
useful discussions.  GF is grateful to V. Vladimirov for
hospitality and support.

\end{document}